\documentclass[12pt]{article}

\usepackage{graphics,epsfig}
\begin{document}

\title{What can we see from Investment Simulation based on Generalized (m,2)-Zipf law?}
\author{Hokky Situngkir\\
(hokky@elka.ee.itb.ac.id)\\
Dept. Computational Sociology\\
Bandung Fe Institute\\
\\
Yohanes Surya\\
(yohaness@centrin.net.id)\\
Surya Research International}
\date{(April 27, 2005)}
\maketitle
%%%%%%%%%%%%%%%%%%%%%
\begin{abstract}

The paper revisits the investment simulation based on strategies
exhibited by Generalized $(m,2)$-Zipf law to present an
interesting characterization of the wildness in financial time
series.  The investigations of dominant strategies on each
specific time series shows that longer words dominant in larger
time scale exhibit shorter dominant ones in smaller time scale and
vice versa. Moreover, denoting the term wildness based on
persistence over short term trend and memory represented by
particular length of words, we can see how wild historical
fluctuations over time series data coped with the Zipf strategies.

\end{abstract}

\begin{center}
$_{\mbox{\footnotesize{\textbf{Keywords:} Generalized (m,2)-Zipf
law, time series, fluctuations, investment.}}}$ \\
\end{center}

%%%%%%%%%%%%%%%%%%%%%%%

\newpage
\baselineskip .25in

\section{Introduction}
In order to choose financial products for investment (e.g.: stock
indexes, future exchange indexes, foreign exchange indexes) one
often aware of how fast a particular financial values change over
time by referring to its historical time series. There are
commonsensical thoughts that some markets are very "wild" for
price varies so fast and the other is not of price moves very
slowly. From the bottom-up perspective, it is apparent that the
price movement will eventually remain the market depth liquidity,
i.e.: order flow necessary to move price by a given amount over
time. In return, other thing that also contributed to this issue
is the 'market climate' surrounding investors or traders
triggering the herding behavior over the market \cite{abmgw}.

Ausloos and Bronlet \cite{ausloossatu} previously have described
the way of texifying the time series data and perform statistical
aspects revealing the Zipf law. In advance, they also propose an
interesting investment strategy based on their findings on
power-laws inspired by the so-called DFA method. The Zipf law of
the financial time series is presented by transforming the ups and
downs of financial index fluctuations into "words" of
\textit{m}-length and \textit{k}-types of "letters"
\cite{ausloosdua}. Obviously, there are a lot of possibilities on
representing the financial indexes into texts, but the rest of the
paper will use the letter \textit{"u"} for higher value over time
respect to the price in the opening and closing session of the
market and \textit{"d"} for downward price movement - whether they
can be weekly, daily, sessional (morning and afternoon market
sessions), and even smaller time intervals, e.g.: hour, minutes,
seconds \cite{vanderwall}. However, concerning the investment
simulation, we will use the smallest interval of hourly
\cite{zipfgw}.

The paper wants to open a new possible analytical door on
financial market by bridging the investment-based paradigm and the
adherence of the generalized Zipf law. We see a possibility on a
new and important way to approach, characterize and extract
information within the time series data by analysis on the
texified fluctuations. The next section of the paper elaborates
the Zipf law in financial data in the way we have the simulated
investment results by using the Generalized Zipf law. This is
followed by discussions on "wildness" of the time series data
concerning the appropriate value of \textit{k} used in the
simulation.

\section{Text from Time Series}
By representing the price fluctuations into, say two alphabets, we
have possible words in the time series data equal to $2^m=2^2$
possible words. They are  \textit{"$uu$"}, \textit{"$dd$"},
\textit{"$ud$"}, and \textit{"$du$"}. This is applied up to
several numbers of possible words. Thus, we sort frequency of
occurrence of the words along our data to have the illustration of
the Zipf plot - rank the sequence started from $R=1$ for the most
frequent words. According to Zipf law, we will have relation of:

\begin{equation}
f \simeq R^{-a}
\end{equation}

In accordance with R/S analysis, Bronlet, et. al.
\cite{ausloosdua} showed the conjecture relation:

\begin{equation}
a = |2H-1|
\end{equation}

It is important to note that the probability of ups \textit{(u)}
and downs \textit{(d)} of price in most of the case is not the
same for every time series data. Here, we found bias:

\begin{equation}
\varepsilon = p(u)-p(d)
\end{equation}

where \textit{p(u)} and \textit{p(d)} each is the probability of
ups and downs in time series data. In this case, we need to put
into account the frequency of expectation \textit{f'}, which
represent random situation (un-correlated). In this case, in our
calculation we use \textit{f'} as the substitution from the
original form of Zipf (\textit{f}) method with coefficient of
exponent \textit{a'}. This mathematically can clearly prevent the
presence of exponent $a=0$ which is possible to occur since those
biases. The value of \textit{f'} can be calculated as

\begin{equation}
f' = p^{m-n}(u).p^{n}(d)
\end{equation}

By now, we can start investment simulation for some financial data
from various markets. This is aimed for us to have basic of
prediction based on scaling character from each financial data in
case of investment. Technically, the question is: what would
likely to happen if we have words sequence with alphabet sequence
length of $(m-1)$?

If we have words sequence of ${c(t-m+1),..., c(t-1), c(t)}$, how
big the probability of \textit{c(t)} to be "up" or "down", with
\textit{c(t)} is showing the characters of time \textit{t}. From
this point, we calculate the prediction based on the words we have
ranked in Zipf plot to find the level of probability of up,
$p_{up}(t)$, that is the occurrence of sequence ${c(t-m+1),..,
c(t-1), "u"}$, and the probability of down, $p_{down}(t)$, or the
occurrence of the sequence ${c(t-m+1),.., c(t-1), "d"}$. It is
suggested \cite{zipfgw} to consider variable of strength parameter
that is how far we can trust a calculation result, which is
through relative probability represented by:

\begin{equation}
D(t) = |\frac{p_{up}(t)-p_{down}(t)}{p_{up}(t)+p_{down}(t)}|
\end{equation}

This \textit{D(t)} parameter shows how big the probability of the
prediction result to have value within interval \textit{$[0..1]$}.
In the simulation, we use parameter \textit{D(t)} as a form of
fraction of how many we \textit{"buy"} or \textit{"sell"} the
index value. As we know, we will pose to \textit{"buy"} when
$p_{up}(t)>p_{down}(t)$ and vice versa.  By the simulation, simply
we can write the outcome of investment after n time as

\begin{equation}
\psi_{total} = \psi_{start} + \sum_{i=2}^{n} (p_i-p_{i-1})D_i
\end{equation}

while in each time-steps we have

\begin{equation}
\psi(t) = \psi(t-1)+[p(t)-p(t-1)]D(t)
\end{equation}

An example of our simulation result is depicted in figure 2
preceding the Zipf law in Jakarta Composite Index (IHSG), DJIA,
and NASDAQ on each daily data in figure 1. Obviously, the figure
shows that the dominant investment strategy in IHSG is
\textit{Zipf (6,2)} while in DJIA and NASDAQ are \textit{Zipf
(4,2)} and \textit{Zipf (5,2)} respectively.

\section{On wildness of financial time series data}
After transforming a series of financial data into text and having
simulation result based on Zipf law in hand, plenty of analytical
thoughts may appear. The notion of relative wildness of financial
data can be treated as the way we cope with the possibilities of
ups and downs of the existing fluctuations. Obviously, when a time
series data can be 'read' well by the Generalized Zipf law, we
have a good view to cope with the wildness of the time series.
Furthermore, the length of the words suitable to gain better along
the certain time window of investment enriches our notion of this
relative 'wildness'. Here we can see how persistent a word along
certain time. For example, if a particular time series data get
better gain while we use the rule of investment of \textit{Zipf
(7,2)} relative to the one of \textit{Zipf (2,2)}, then we can say
that the wildness of the one with \textit{Zipf (7,2)} is wilder
than the other one. In other words, we can say that concerning the
short term trend persistence, time series represented by $7$ words
in investment is less wild relative to the one with $2$ words.

In advance, we can also assume that the persistence over
particular words as a memory effect of time series data. In this
perspective, apparently a less wild time series has longer memory
since the persistent trend is longer. This is very interesting
when we try to see the effect in several specific financial data,
e.g.: financial indexes and foreign exchange rates.

In figure $3$ and $4$, we show the result of our investigation in
daily Yen/USD and Euro/USD data showing good gain in \textit{Zipf
(3,2)} and \textit{Zipf (2,2)} respectively. Comparing to the
previous one in figure (2), we can say relatively that both
foreign exchange rates exhibit wilder fluctuations regarding
persistence over trends and memory represented by the specific
words. Certainly, we can understand this since foreign exchange
involves more traders with higher sensitivity to fundamental issue
relative to the market indexes.

The later question is whether or not the better strategies in a
certain financial products are persistent over different time
scales. We do the similar analysis to HangSeng Index, NIKKEI225
index, and GPB/USD rate over different time orders. Figure (9) and
(12) shows our results with HangSeng and NIKKEI225 indexes on
respective time series data depicted in figure (7) and (10).
Interestingly, a wilder time series with shorter dominant words in
small time scale have longer words of dominant strategies in the
larger time scale and vice versa. This is showed by HangSeng Index
whose dominant strategy of \textit{Zipf (7,2)} in daily data
shrinks upto \textit{Zipf (4,2)} in sessional data and
\textit{Zipf (3,2)} in hourly in our investment simulation. In
return, the dominant \textit{Zipf (3,2)} in daily NIKKEI225
investment simulation exhibits longer \textit{Zipf (6,2)} in
sessional and hourly trading. This is very interesting while
GBP/USD rate shows similar situations of \textit{Zipf (7,2)} in
daily data and \textit{Zipf (8,2)} in hourly simulations.

Intuitively, by referring to the wildness of a time series, we
could say that there could be a pattern presented here. The time
series shows less 'wildness' in a certain time scale could have
wilder in bigger or narrower time scale. This opens further
investigations in spite of the fact that the Generalized Zipf
analysis could yield a good prediction of time series
fluctuations.

\section{Concluding Remarks and Further Works}
We show the simulation result of investment by using the
strategies laid upon our understanding of Zipf law in texified
financial fluctuation. We evaluate the different strategies
regarding the length of words used in each simulation and discover
interesting properties over different time series data and
different time scale. In our simulations we can see that longer
words dominant in larger time scale exhibit shorter dominant ones
in smaller time scale and vice versa. In the other hand, we
present that some financial indexes are wilder than other respect
to the persistence of short term trend (represented by the words
or sequence of letters) and memory over fluctuation patterns.

We expect to have more properties on doing several further
investigations by involving more letter ($k>2$) and by trying to
relate the findings with terminologies often used to analyze the
memory, distributions, and correlations of the time series. Here,
notwithstanding, we have seen some possible and interesting
findings of the textual analysis upon texified time series data
more than the advantage of the analysis as a good and promising
prediction method.

\section{Acknowledgement}
Authors thank M. Ausloos for some important literatures, Yohanis
and Jackson Silaban for financial data, Tiktik Dewi Sartika for
some typeset corrections, and BFI colleagues for discussions.
Noone but authors are responsible for possible errors and
omissions.

%%%%%%%%%%%%%%%%%

%%%%%%%%%%%%%%%%%%%

\begin{figure}
\epsfxsize=\columnwidth
\begin{center}
\centerline{\epsfbox{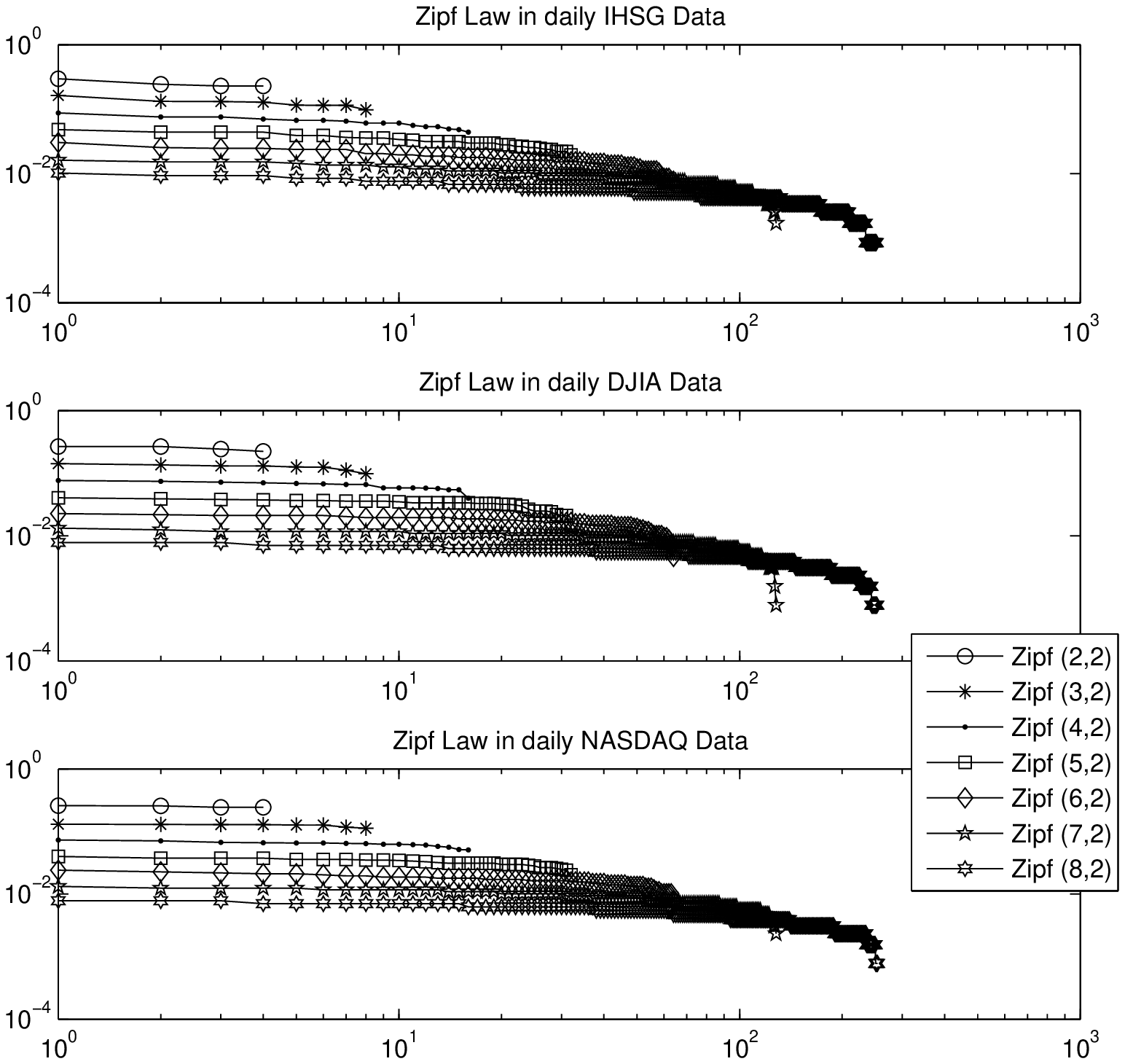}} \caption{Zipf law for various
length of words in Jakarta Composite Index (IHSG), DJIA, and
NASDAQ.  The data is in interval: IHSG (January 4, 2000 - November
2, 2004), DJIA (January 3, 2000 - February 14, 2005), and NASDAQ
(January 3, 2000 - February 14, 2005)}
\end{center}
\end{figure}

\begin{figure}
\epsfxsize=\columnwidth
\begin{center}
\centerline{\epsfbox{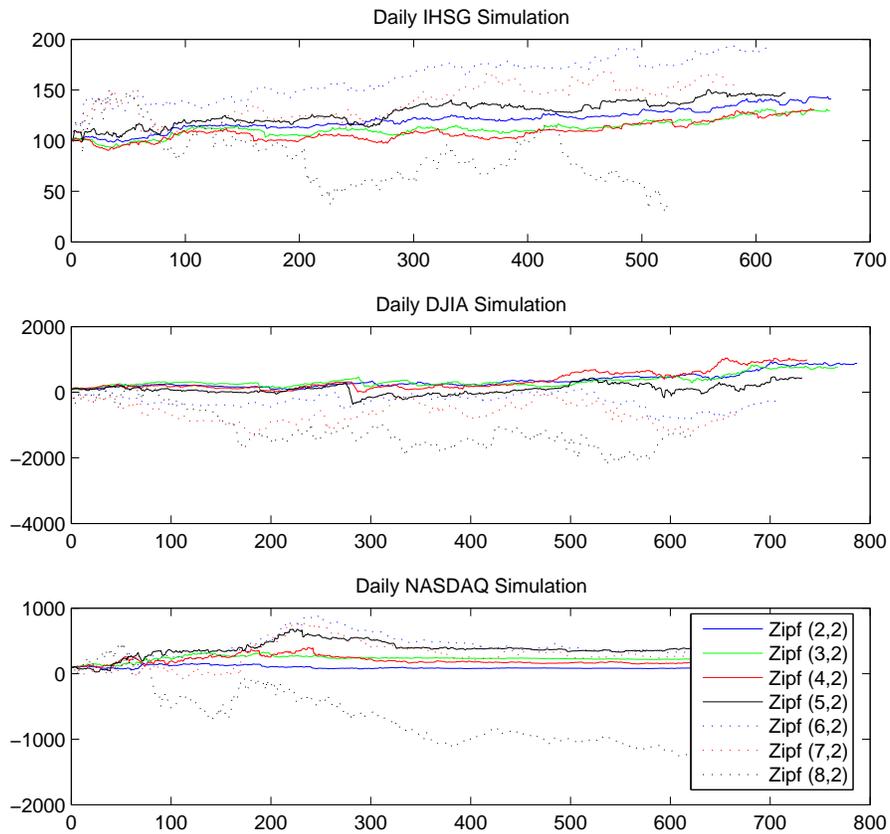}} \caption{Investment Simulation
Result showing the gained point by using the series of Jakarta
Composite Index (IHSG), DJIA, and NASDAQ }
\end{center}
\end{figure}

\begin{figure}
\epsfxsize=\columnwidth
\begin{center}
\centerline{\epsfbox{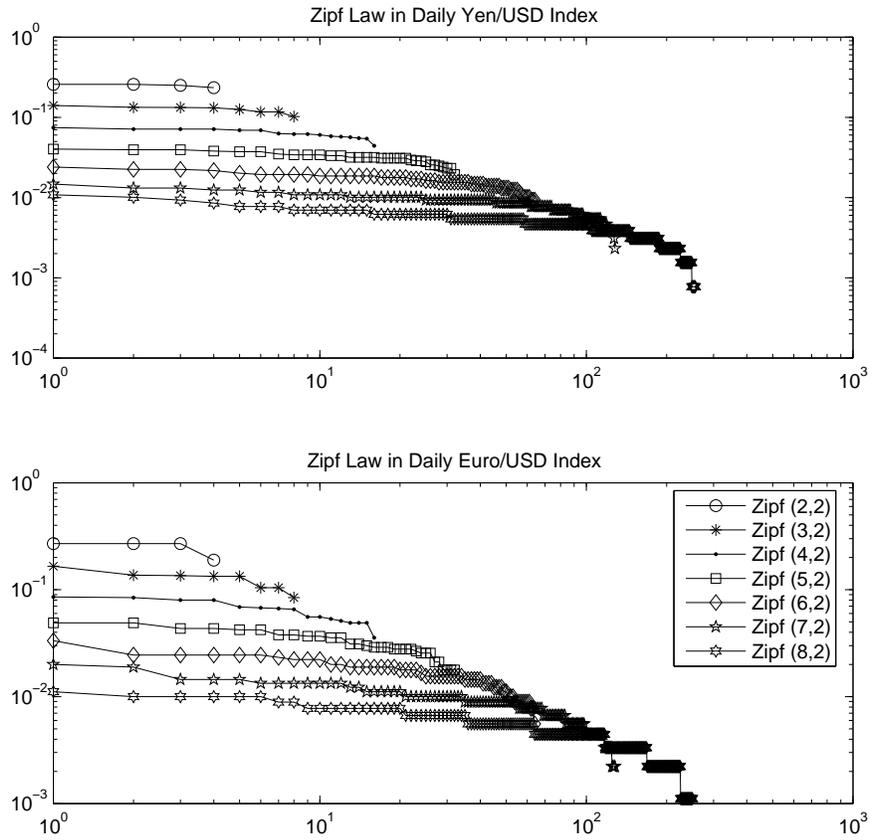}} \caption{Zipf Law in Yen/USD
rate and Euro/USD Rate. The data is in interval: Yen/USD (January
3, 2000 - December 14, 2004) and Euro/USD (July 9, 2001 - December
17 2004) }
\end{center}
\end{figure}

\begin{figure}
\epsfxsize=\columnwidth
\begin{center}
\centerline{\epsfbox{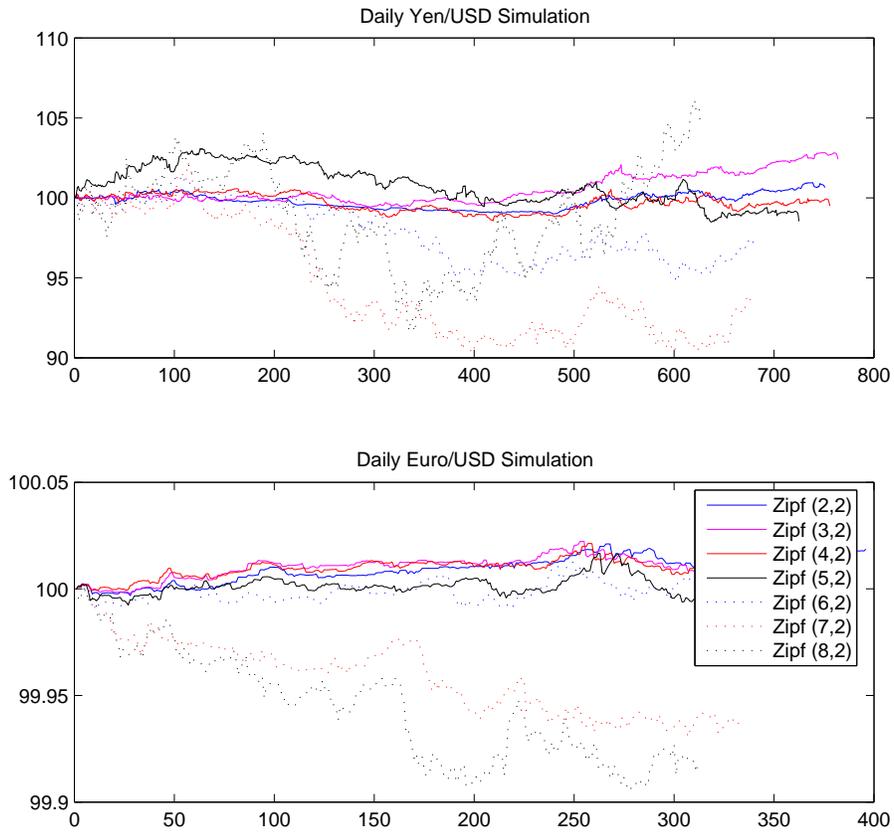}} \caption{Investment Simulation
Result showing the gained point by using the series of Yen/USD and
Euro/USD rate. }
\end{center}
\end{figure}

\begin{figure}
\epsfxsize=\columnwidth
\begin{center}
\centerline{\epsfbox{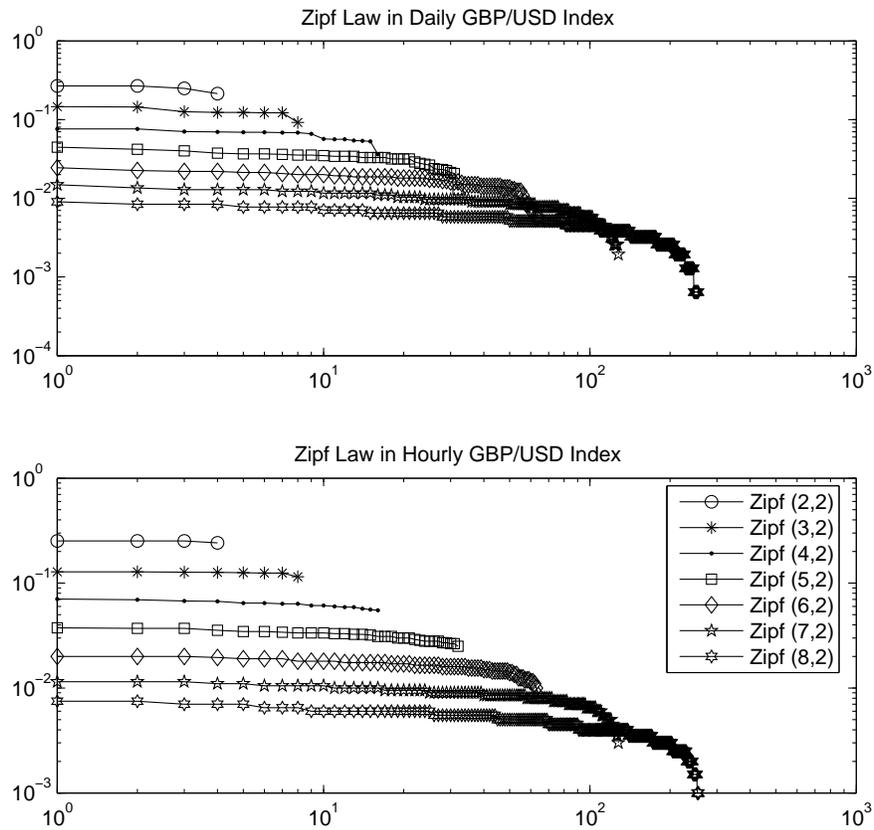}} \caption{Zipf Law in GBP/USD
Rate for daily (January, 3, 2000 - December 14 2004) and hourly
data (22:00 September 30, 2004 - 12:00 January 24 2005. }
\end{center}
\end{figure}

\begin{figure}
\epsfxsize=\columnwidth
\begin{center}
\centerline{\epsfbox{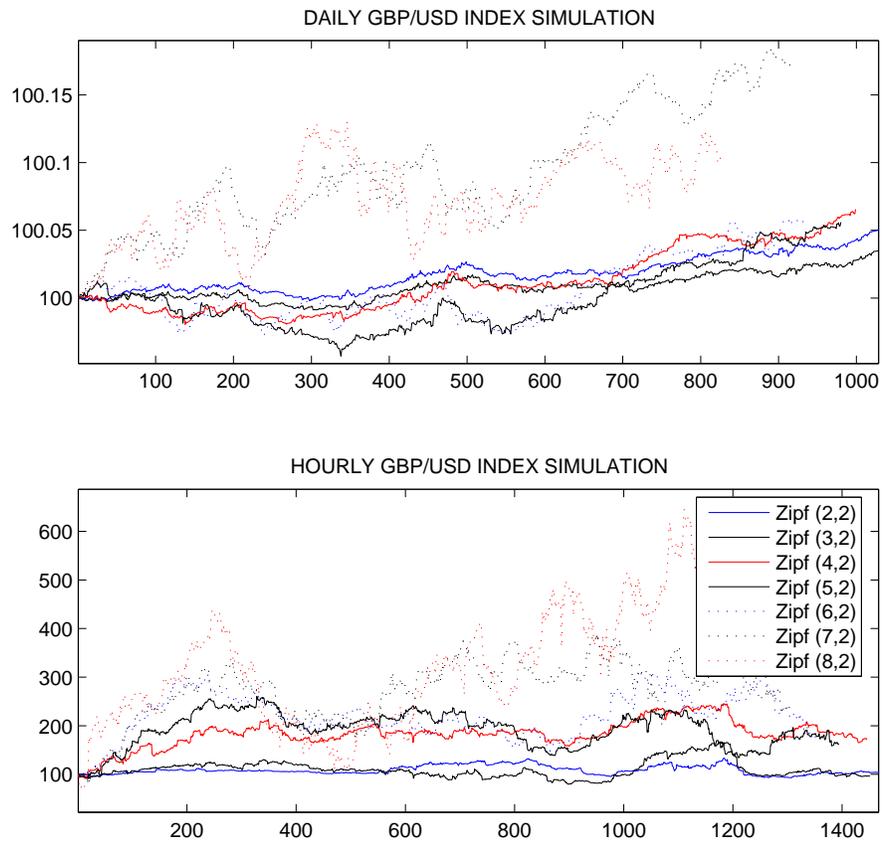}} \caption{Investment Simulation
Result for daily and hourly trading. }
\end{center}
\end{figure}

\begin{figure}
\epsfxsize=\columnwidth
\begin{center}
\centerline{\epsfbox{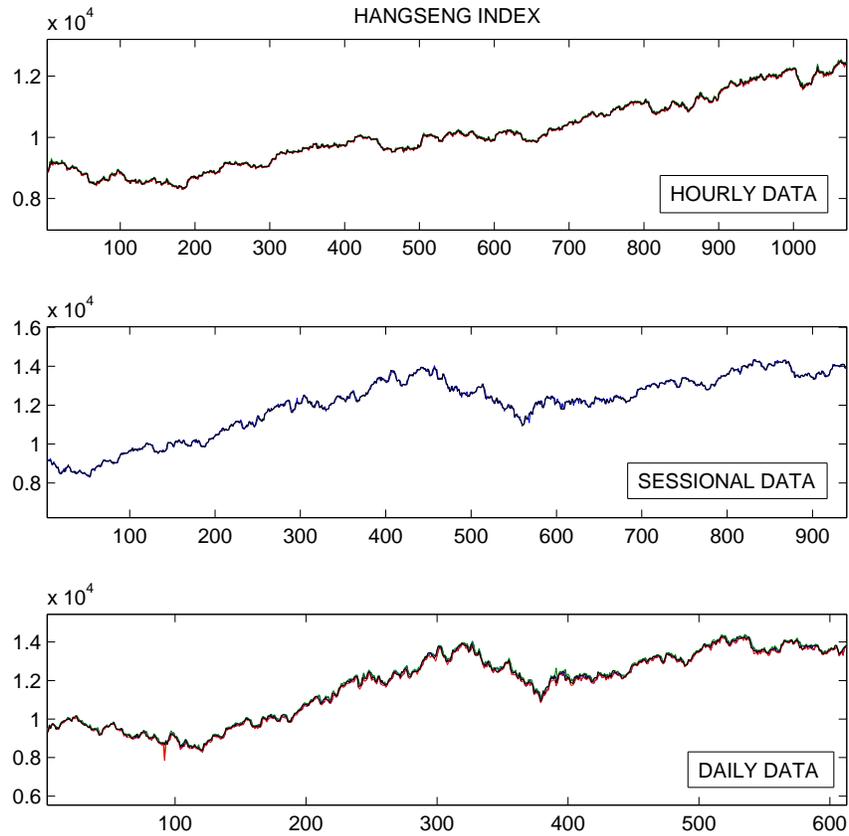}} \caption{HangSeng Index used
in investment simulation for daily (October 30, 2002 - April 26,
2005), sessional (March 19, 2003 - September 3, 2004), and hourly
(10:00 March 19, 2003 - 16:00 August 28, 2003) data respectively.
}
\end{center}
\end{figure}

\begin{figure}
\epsfxsize=\columnwidth
\begin{center}
\centerline{\epsfbox{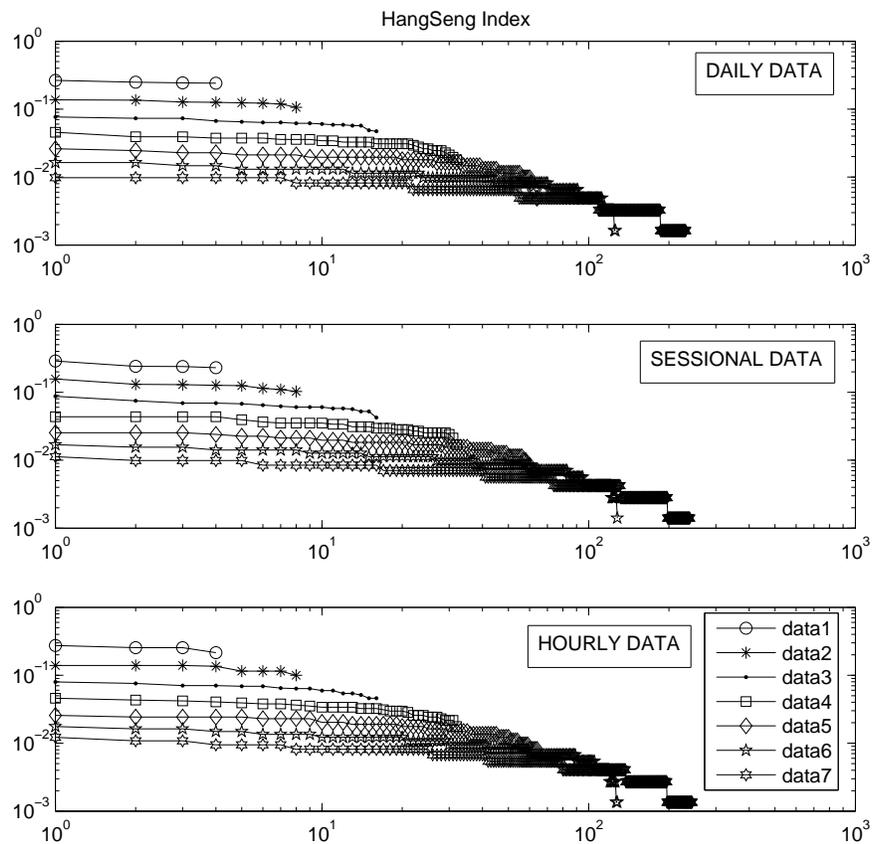}} \caption{Zipf Law in texified
HangSeng Index. }
\end{center}
\end{figure}

\begin{figure}
\epsfxsize=\columnwidth
\begin{center}
\centerline{\epsfbox{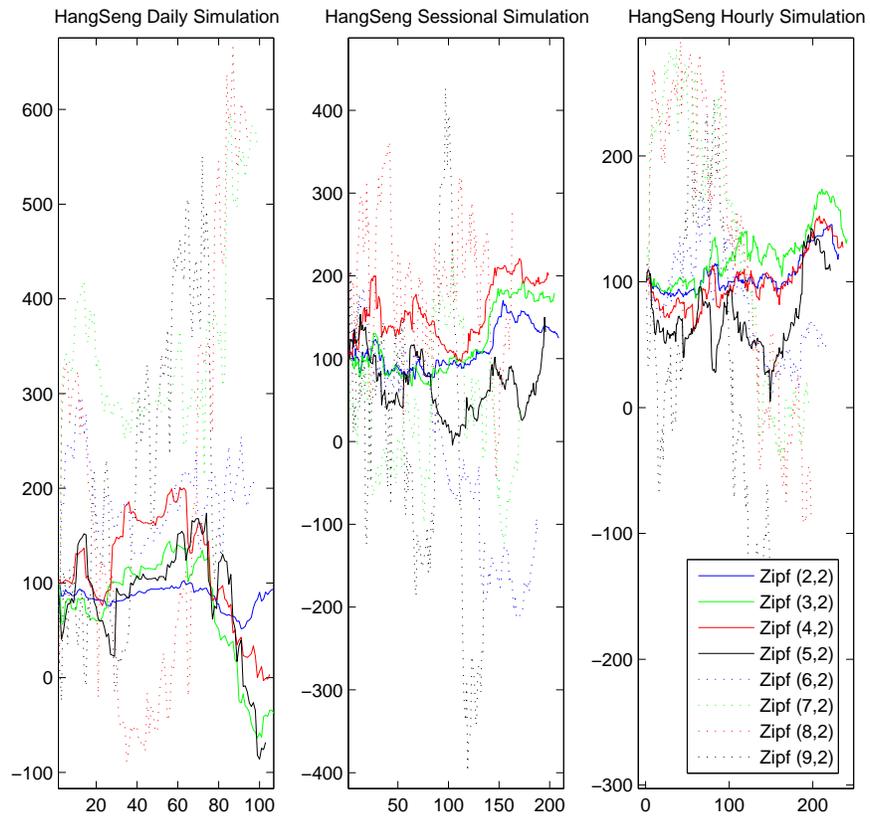}} \caption{Simulation result
comparing investment by respective (m,2)-Zipf law in daily,
sessional, and hourly HangSeng market. }
\end{center}
\end{figure}

\begin{figure}
\epsfxsize=\columnwidth
\begin{center}
\centerline{\epsfbox{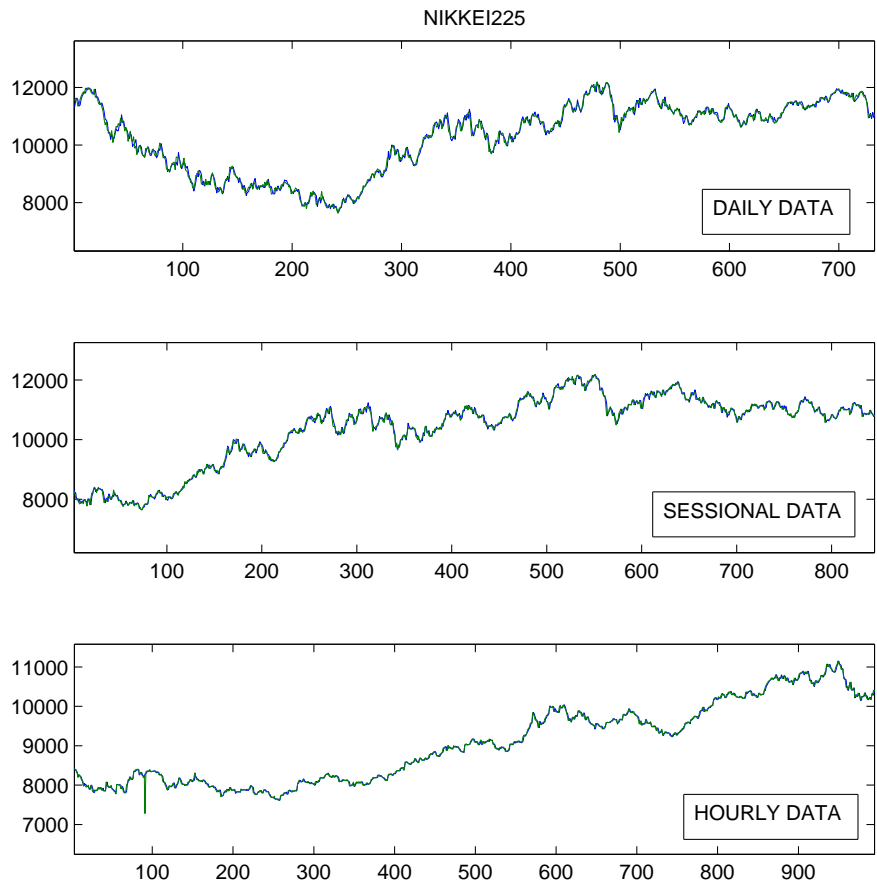}} \caption{NIKKEI225 Index used
in investment simulation for daily (May 8, 2002 - April 26, 2005),
sessional (March 6, 2003 - December 2, 2004), and hourly (8:00
March 6, 2003 - 12:00 October 1, 2003) data respectively. }
\end{center}
\end{figure}

\begin{figure}
\epsfxsize=\columnwidth
\begin{center}
\centerline{\epsfbox{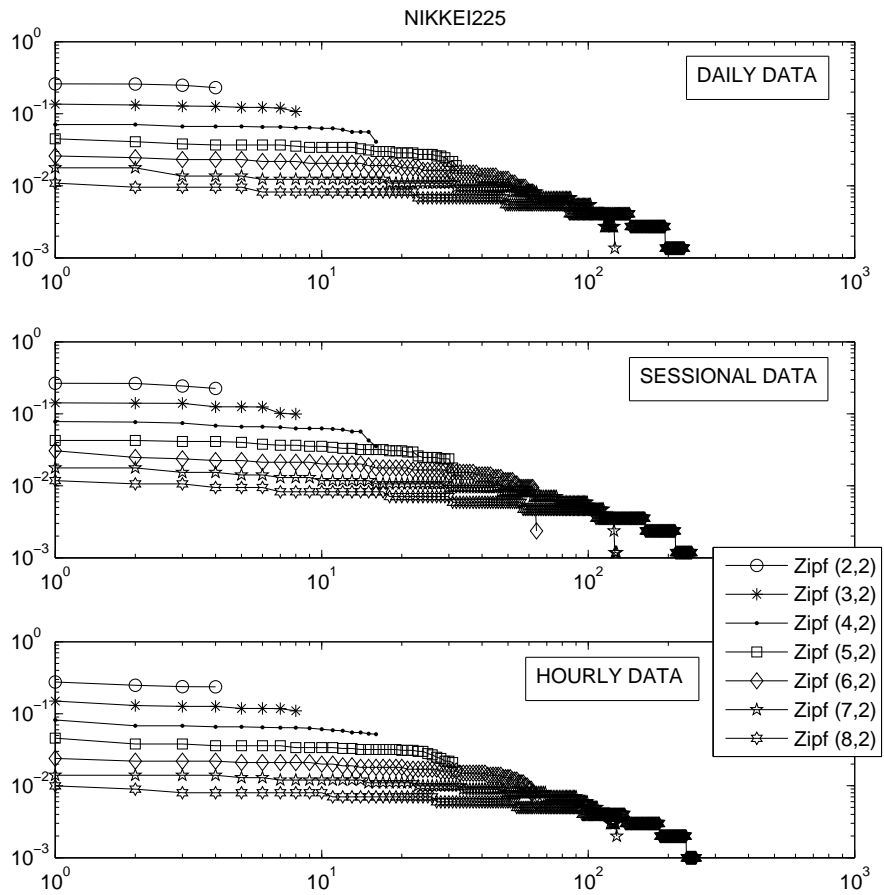}} \caption{Zipf Law in texified
NIKKEI225 Index. }
\end{center}
\end{figure}

\begin{figure}
\epsfxsize=\columnwidth
\begin{center}
\centerline{\epsfbox{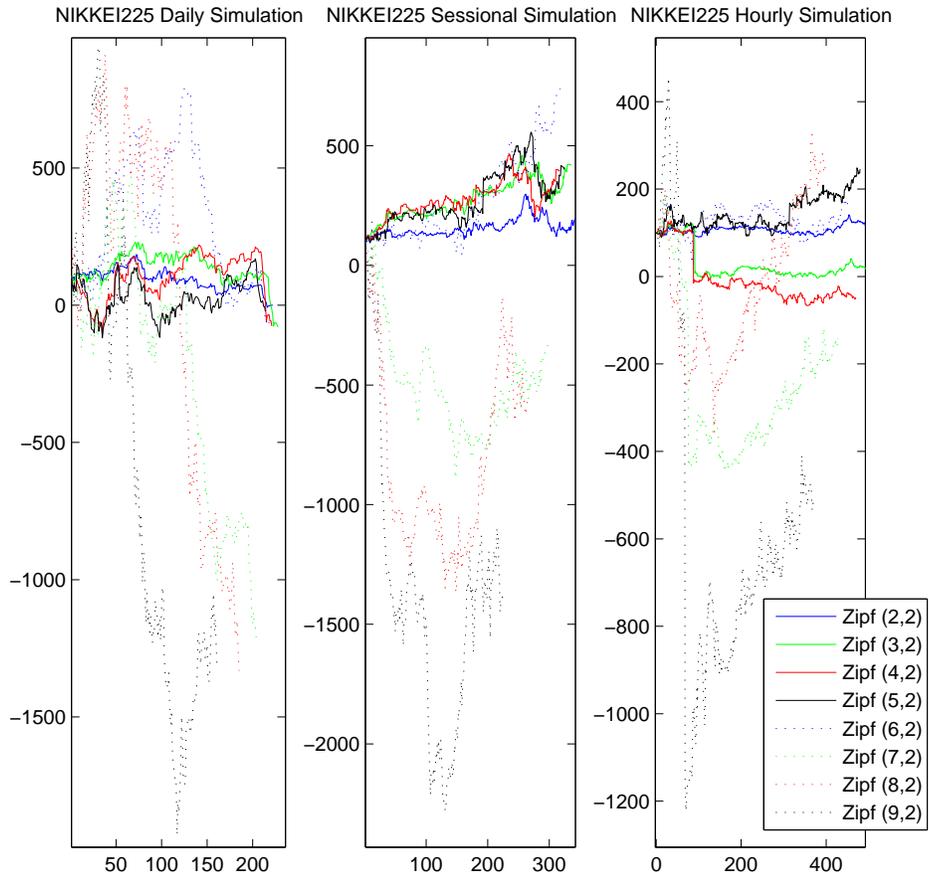}} \caption{Simulation result
comparing investment by respective (m,2)-Zipf law in daily,
sessional, and hourly NIKKEI225 market. }
\end{center}
\end{figure}

%%%%%%%%%%%%%%%%%
\end{document}